\documentclass[10pt,a4,conference]{IEEEtran}
\IEEEoverridecommandlockouts
\usepackage{cite}
\usepackage{amsmath,amssymb,amsfonts}
\usepackage{algorithmic}
\usepackage{graphicx}
\usepackage{textcomp}
\usepackage{xcolor}
\usepackage{caption}
\usepackage{subcaption}
\usepackage{makecell}
\usepackage{multirow}
\usepackage{comment}
\usepackage{balance}

\usepackage{caption}
\usepackage{colortbl}
\usepackage{color}
\definecolor{light_grey}{rgb}{0.8,0.8,0.8}
\usepackage{dblfloatfix}

\usepackage[
top    = 0.75in,
bottom = 1.05in,
left   = 0.673in,
right  = 0.673in]{geometry}

\DeclareMathOperator*{\argmax}{arg\,max}

\makeatletter 
\newcommand{\linebreakand}{%
  \end{@IEEEauthorhalign}
  \hfill\mbox{}\par
  \mbox{}\hfill\begin{@IEEEauthorhalign}
}
\makeatother 

\def\BibTeX{{\rm B\kern-.05em{\sc i\kern-.025em b}\kern-.08em
    T\kern-.1667em\lower.7ex\hbox{E}\kern-.125emX}}

\IEEEoverridecommandlockouts
\usepackage{tikz}
\usepackage{textcomp}
\usepackage[hidelinks]{hyperref}
\usepackage{lipsum}
\newcommand\copyrighttext{%
  \footnotesize \textcopyright 2025 IEEE. Personal use of this material is permitted.  Permission from IEEE must be obtained for all other uses, in any current or future  media, including reprinting/republishing this material for advertising or promotional  purposes, creating new collective works, for resale or redistribution to servers or  lists, or reuse of any copyrighted component of this work in other works.
  DOI: \href{<http://tex.stackexchange.com>}{10.1109/GLOBECOM59602.2025.11432718}
  }
\newcommand\copyrightnotice{%
\begin{tikzpicture}[remember picture,overlay]
\node[anchor=south,yshift=10pt] at (current page.south) {\fbox{\parbox{\dimexpr\textwidth-\fboxsep-\fboxrule\relax}{\copyrighttext}}};
\end{tikzpicture}%
}

\begin{document}

\title{Angle diversity receiver as a key enabler for reliable ORIS-based Visible Light Communication

\thanks{Borja Genoves Guzman has received funding from the European Union under the Marie Skłodowska-Curie grant agreement No 101061853, and from Ramón y Cajal grant RYC2023-042518-I, funded by MCIU/AEI/10.13039/501100011033 and FSE+. M. Morales Cespedes has received funding from Ramón y Cajal grant RYC2022-036053-I, MICIU/AEI/10.13039/501100011033 and FSE+. This work has been partially funded by project PID2023-147305OB-C31 (SOFIA-AIR) by MCIN/ AEI/10.13039/501100011033/ERDF, UE, by project TUCAN6-CM (TEC-2024/COM-460), funded by CM (ORDEN 5696/2024), by project PID2024-156038OA-I00 (HOLIFI6G) by MICIU /AEI /10.13039/501100011033 / FEDER, UE, and by ViDiT project TED2021-129869B-I00 funded by MICIU/AEI/ 10.13039/501100011033 and ``EU NextGeneration EU/PRTR''.}
}

\author{
\IEEEauthorblockN{Borja Genoves Guzman\IEEEauthorrefmark{1}, M\'aximo Morales-C\'espedes\IEEEauthorrefmark{1}
Ana Garc\'ia Armada\IEEEauthorrefmark{1}, and
Maïté Brandt-Pearce\IEEEauthorrefmark{2}}
\IEEEauthorblockA{\IEEEauthorrefmark{1}Signal Theory and Communications Dept., University Carlos III of Madrid, Leganes, Madrid 28911 Spain}
\IEEEauthorblockA{\IEEEauthorrefmark{2}Electrical and Computer Engineering Dept.,
University of Virginia, Charlottesville, VA 22904 USA}
\IEEEauthorblockA{E-mails: bgenoves@ing.uc3m.es, mmcesped@ing.uc3m.es, anagar@ing.uc3m.es, mb-p@virginia.edu}
}

\maketitle

\copyrightnotice 
\vspace{-\baselineskip}

\begin{abstract}
Visible Light Communication (VLC) offers a promising solution to satisfy the increasing demand for wireless data. However, link blockages remain a significant challenge. This paper addresses this issue by investigating the combined use of angle diversity receivers (ADRs) and optical reconfigurable intelligent surfaces (ORISs) in multiuser VLC systems. We consider ORIS elements as small movable mirrors. We demonstrate the complementarity of ADR and ORIS in mitigating link blockages, as well as the advantages of using a larger number of ORIS elements due to the increased field-of-view (FoV) at the receiver enabled by the ADR. An optimization algorithm is proposed to maximize the minimum signal-to-noise power ratio (SNR) to deploy a fair communication network. Numerical results show that integrating ADR and ORIS significantly enhances VLC communication performance, achieving an SNR gain of up to 30\,dB compared to a system without ORIS, and mitigating communication outages produced by link blockages or out-of-FoV received signals. We also prove that an ADR with a single tier of photodiodes is sufficient to complement ORIS-assisted VLC.
\end{abstract}


\vspace{-1mm}
\section{Introduction}\label{sec:Intro}
\vspace{-1mm}

Visible light communication (VLC) has emerged as a promising technology to complement traditional radio frequency systems for indoor wireless communications.  VLC leverages the widespread deployment of light-emitting diodes (LEDs) to transmit data over the untapped and unregulated visible light spectrum. The transmission and reception are based on intensity modulation and direct detection (IM/DD), respectively. However, these wavelengths are susceptible to significant path loss, particularly when opaque objects obstruct the direct path between transmitter and receiver, causing a so-called outage~\cite{LightsShadows}.

To mitigate this issue, optical reconfigurable intelligent surfaces (ORISs) have recently been proposed as a means of establishing robust, non-line-of-sight (NLoS) paths~\cite{MirrorVSMetasurface}.  In~\cite{MirrorVSMetasurface}, the authors considered both mirrors and metasurfaces as potential ORIS elements, concluding that mirrors may offer superior communication performance.  This paper adopts ORISs composed of small mirrors mounted on an electromechanical system, enabling precise control over the direction of light reflection.  Unlike diffuse and weak reflections from walls, ORIS reflections are controlled, powerful, and directed towards the user's location.

\begin{figure}[t]
\centering
\includegraphics[width=\columnwidth]{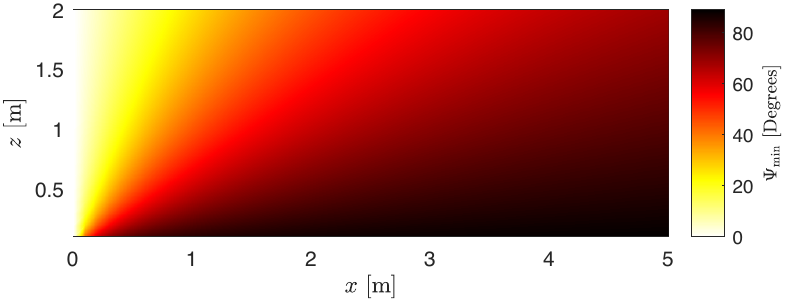}
        \caption{Minimum half FoV angle required at each user position.}
\label{fig:MinFoV}
\vspace{-5mm}
\end{figure}

ORIS-assisted VLC systems offer spatial diversity, which can potentially improve several aspects of communication, including data rates~\cite{SumRateMirrors}, spectral efficiency~\cite{JointResourceMirrors}, secrecy rates~\cite{ORISSecrecy}, illumination uniformity~\cite{MirrorVLC}, and outage probability~\cite{Globecom2023,guzman2024resource,guzman2024usingcurvedmirrorsdecrease,guzman2025orisallocationminimizeoutage}.  Specifically, the outage performance of ORIS-assisted VLC in a multiuser scenario is evaluated in~\cite{guzman2025orisallocationminimizeoutage}, demonstrating the technology's potential to overcome link blockages.  However, ORIS-assisted VLC systems typically require a large field-of-view (FoV) at the receiver to capture signals from ORIS elements positioned on walls.  Achieving this large FoV with a single photodiode is challenging in practice. The use of concentrators may solve this issue at the cost of losing receiver gain~\cite{VLC_Concentrators}. 
Angle diversity receivers (ADRs) offer an alternative solution. ADRs use multiple small receivers, each with a narrow FoV, oriented in different directions to emulate a single receiver with a large FoV~\cite{VLC_ADR1,VLC_ADR2}.  This approach enables the reception of signals from multiple angles without sacrificing receiver gain and facilitate the use of advanced signal processing techniques to mitigate interference~\cite{InterferenceVLC}.  However, fully exploiting the performance of ADRs requires channel diversity, which can be difficult to achieve in typical VLC scenarios.  Because ORIS elements can provide diversity gain and require large receiver FoVs, a combination of ADR and ORIS is promising in VLC scenarios.

To the best of the authors' knowledge, no prior works have studied the joint performance of ORIS and ADR. This study works at their intersection and evaluates the potential of ADR and ORIS complementarity. First, we provide a system model in which both ORIS elements and ADRs are introduced. Then, we propose an optimization problem to maximize the minimum signal-to-noise power ratio (SNR) and then optimally associate ORIS elements with users and LEDs, while considering the ADR arrangement. Finally, we provide simulation results and show that the complementary use of ORIS and ADR leads to a considerable enhancement of the VLC performance in terms of SNR and sum rate. 

\begin{figure}[t]
\centering
\includegraphics[width=0.95\columnwidth]{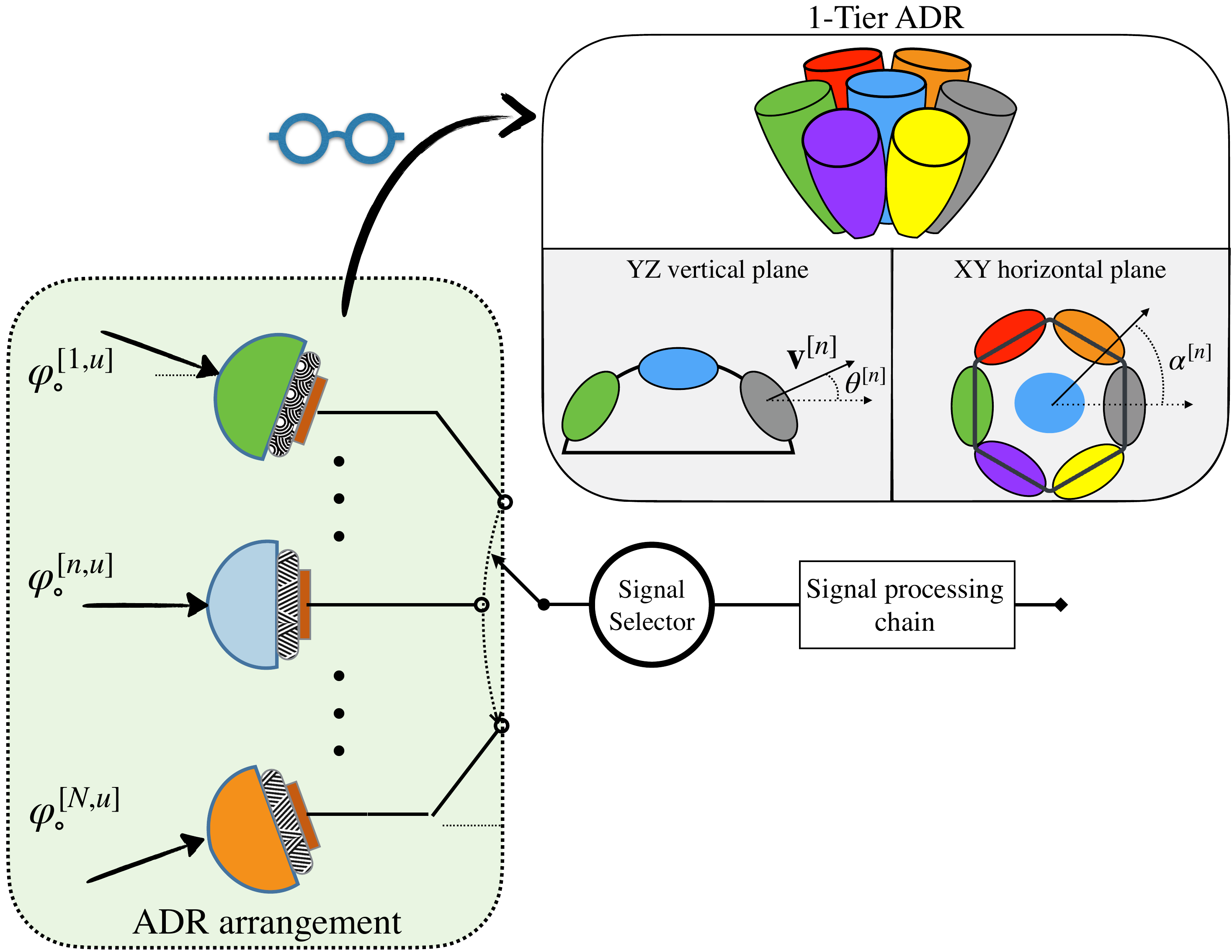}
        \caption{ADR assuming a reconfigurable photodetector architecture. The set of photodiodes follows an angular diversity arrangement connected to a single signal processing chain through a selector. For the sake of generality, $\varphi^{[n,u]}_{\circ}$ denotes the incidence angle at photodiode $n$ of user $u$ from the source $\circ$, either an AP or an ORIS element. The angular distribution of the azimuthal and elevation angles for a 1-Tier configuration is highlighted. }
\label{fig:reconfigurable}
\vspace{-6mm}
\end{figure}

\section{System Model}\label{sec:SO}

We consider a VLC scenario composed of $L$, $l = \{1, \dots, L\}$ LED-based optical access points (APs), transmitting to $U$, $u = \{1, \dots, U\}$ users distributed in the room. Assuming that the user receiver is looking upwards, and according to eq.\,(11) in~\cite{guzman2024resource}, the minimum half FoV required ($\Psi_{\rm min}$) to receive a signal from an ORIS located at a horizontal distance $x$ and a vertical distance $z$ with respect to a user, is given by
\vspace{-1mm}
\begin{IEEEeqnarray}{rCl}
\Psi_{\rm min} & = & \arctan{\frac{x}{z}},
\vspace{-1mm}
\label{eq:MaxDistRIS}
\end{IEEEeqnarray}
which is represented in Fig.\,\ref{fig:MinFoV}. Notice that the required minimum half FoV is mostly greater than 60$^\circ$, results which are difficult to achieve considering a single photodiode. Motivated by this issue, in this paper, we equip each user with an ADR composed of $N$, $n= \{1,\dots, N\}$ photodiodes, increasing the effective FoV of the receiver. Specifically, the $N$ photodiodes are deployed following an angular arrangement as the one represented in Fig.\,\ref{fig:reconfigurable}. That is, the pointing vector of photodiode $n$ can be written as 
\vspace{-1mm}
\begin{equation}
\hat{\mathbf{v}}^{[n]} {=} 
{\begin{bmatrix}
\cos\hspace{-1mm}\left(\alpha^{[n]}\right)\sin\hspace{-1mm}\left(\theta^{[n]}\right) & \sin\hspace{-1mm}\left(\alpha^{[n]}\right)\sin\hspace{-1mm}\left(\theta^{[n]}\right) &
\cos\hspace{-1mm}\left(\theta^{[n]}\right)
\end{bmatrix}},
\end{equation}
where $\alpha^{[n]}$ and $\theta^{[n]}$ are the azimuthal and elevation angles of photodiode $n$, respectively. Based on this approach, we consider a reconfigurable photodetector configuration in which all photodiodes are connected to a single signal processing chain through a selector, as shown in Fig.~\ref{fig:reconfigurable}. In this way, the user will connect the best-performing photodiode to a single processing chain, according to the environmental conditions at each moment. Due to their small dimensions, LEDs and photodiodes are modeled as point sources and detectors, respectively. We also assume that all photodiodes of the same user are collocated.


\begin{figure}[t]
     \centering
     \begin{subfigure}[t]{0.37\columnwidth}
         \centering
         \includegraphics[width=\textwidth]{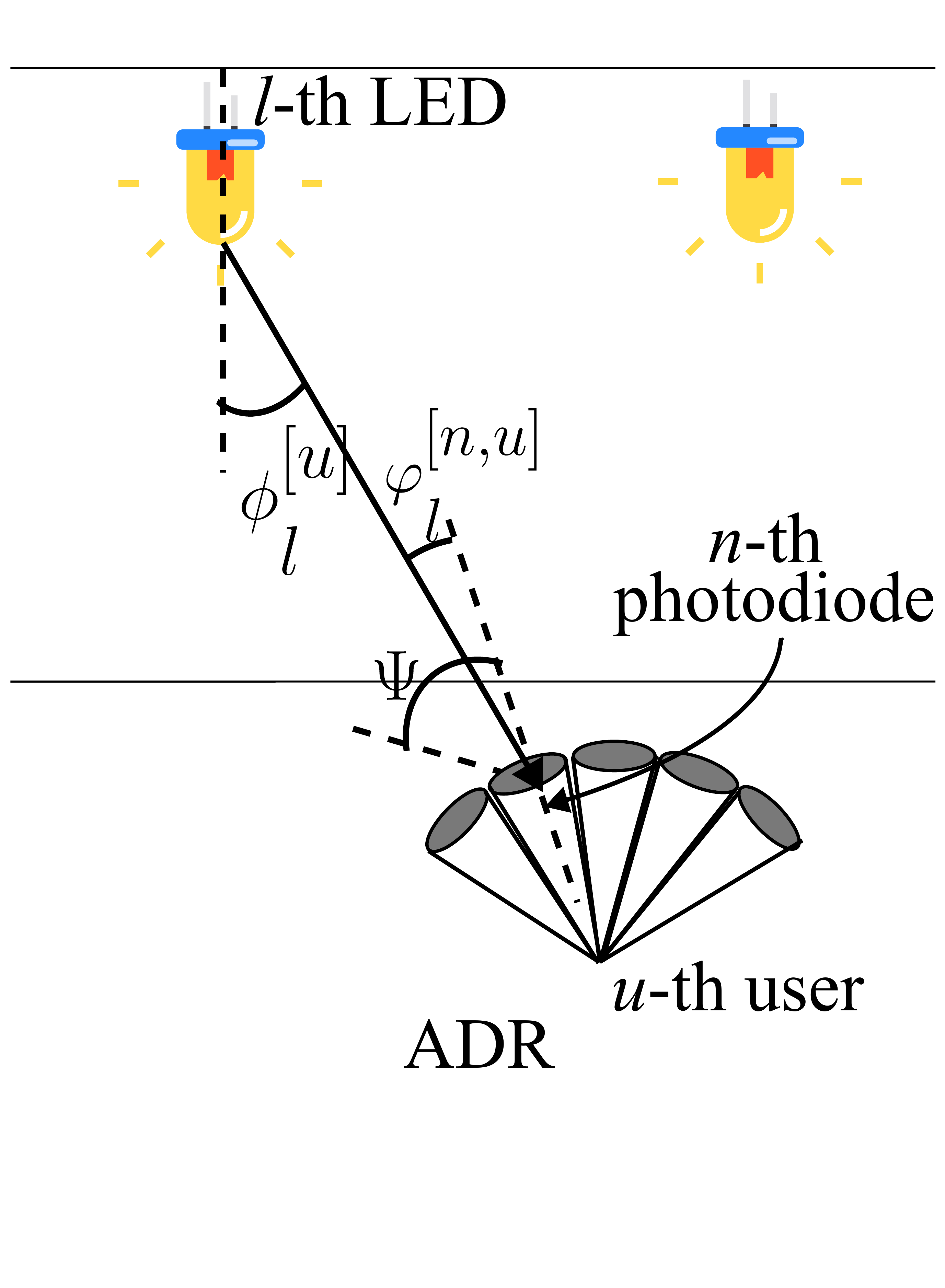}
         \caption{LoS}
         \label{fig:ScenarioLoS}
     \end{subfigure}
     \hfill
     \begin{subfigure}[t]{0.57\columnwidth}
         \centering
         \includegraphics[width=\textwidth]{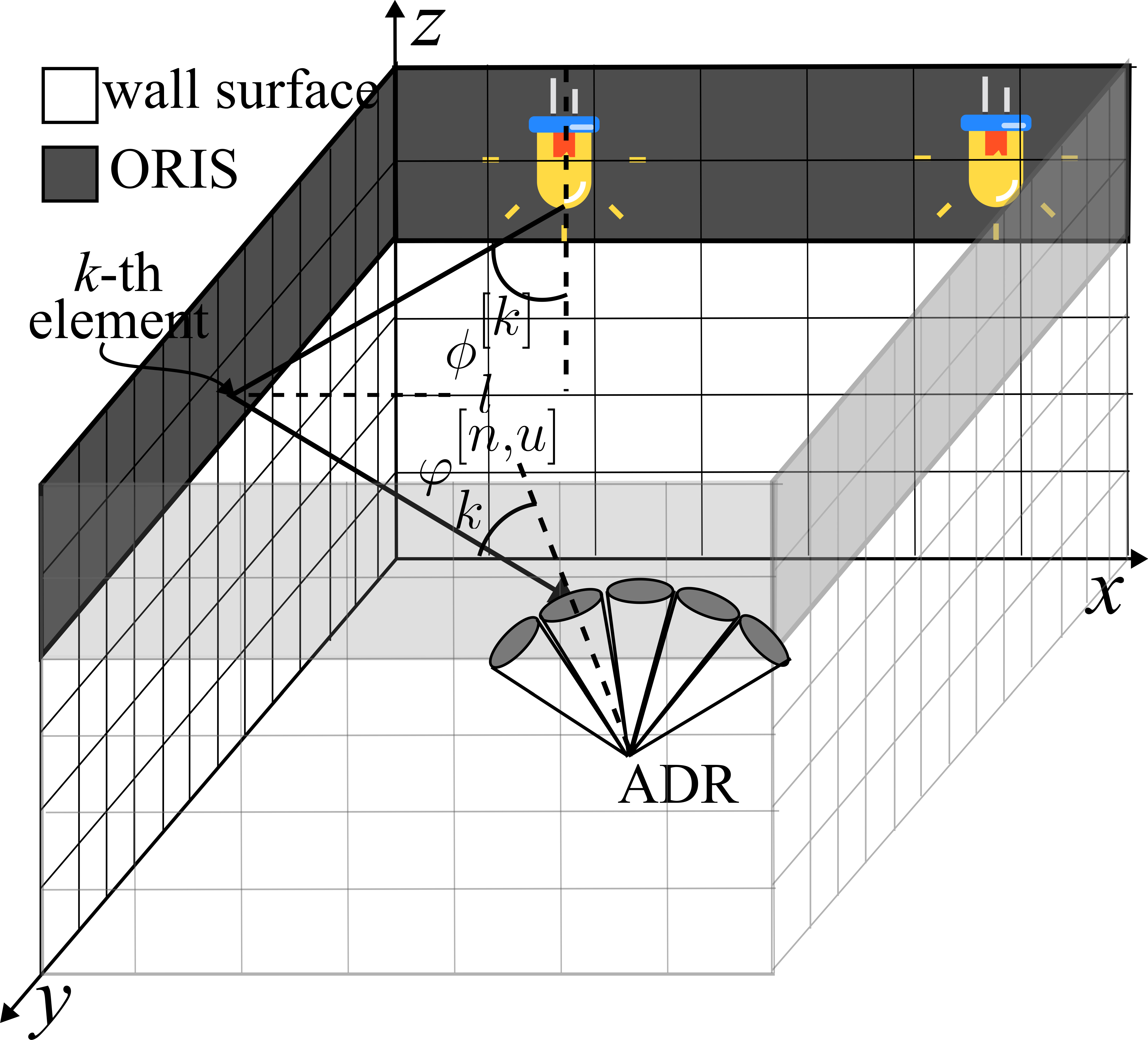}
         \caption{NLoS}
         \label{fig:ScenarioNLoS}
     \end{subfigure}
        \caption{System model: 3D illustration of LoS and NLoS propagation. ORIS (mirrors) are placed forming a crown molding in the room.}        
        \label{fig:SystemModel}
        \vspace{-6mm}
\end{figure}

The communication performance relies on line-of-sight (LoS) and NLoS links defined as follows. The LoS channel gain from AP $l$ to photodiode $n$ of user $u$, denoted by $H_{l, \rm{LoS}}^{{[n,u]}}$, is modeled using the Lambertian emission pattern~\cite{VLCLoSChannel},
\vspace{-1mm}
\begin{equation}
\label{eq:ChannelModelLoS}
H_{l, \rm{LoS}}^{{[n,u]}} {=}\hspace{-1mm}
\begin {cases}
\hspace{-1mm}I_{l}^{[u]}\frac{{\left( {m + 1} \right) \cdot {A_{{\rm{PD}}}}}}{{2\pi \left(d_{l}^{[u]}\right)^2}}{{\cos }^m}\hspace{-1mm}\left( \hspace{-0.5mm} {{\phi_{l}^{[u]}}} \right)\hspace{-1mm}\cos \hspace{-1mm}\left( \hspace{-0.5mm}{{\varphi_{l}^{[n,u]}}} \right) & 0 {\le} {\varphi_{l}^{[n,u]}} {\le} {\Psi}\\ 
\hspace{-1mm} 0 & \rm{otherwise},
\end{cases}
\end{equation}
where $m=-1/\log_2\left(\cos\left(\phi_{1/2}\right)\right)$ represents the Lambertian order, characterizing the LED's radiation profile, which is determined by its half-power semi-angle $\phi_{1/2}$. The active area of the photodiode is given by $A_{{\rm PD}}$, and $d_{l}^{[u]}$ stands for the Euclidean distance between AP $l$ and user $u$. The irradiance angle and the incidence angle are denoted by $\phi_{l}^{[u]}$ and $\varphi_{l}^{[n,u]}$, respectively, as illustrated in Fig.\,\ref{fig:ScenarioLoS}.  The indicator variable $I_{l}^{[u]}$ is binary, taking the value 1 if a LoS path exists between AP $l$ and user $u$, and 0 if it is blocked. The FoV is denoted by $\Psi$, and it is the same for all photodiodes.

The NLoS channel is formed by reflections from ORIS and wall elements. In this paper, we consider each wall to be segmented into a grid comprising $K$, $k = \{1, \dots, K\}$ ORIS elements, and $W$, $w = \{1, \dots, W\}$ wall elements, resulting in a total of $K+W$ elements per wall.  Based on the findings in~\cite{guzman2024resource}, and as it can also be seen in Fig.\,\ref{fig:MinFoV}, the optimal placement for ORIS is at the top of each wall.  Consequently, this paper adopts a ``crown molding'' configuration as in~\cite{guzman2025orisallocationminimizeoutage}, where ORIS elements are positioned along the upper sections of each wall, as depicted in Fig.\,\ref{fig:ScenarioNLoS}.  Each ORIS element is mounted on an electromechanical system, enabling optimal orientation of the mirrors to direct reflections toward the receiver.  We assume that blockage between reflectors is negligible.  

The NLoS channel gain from an ORIS element $k$, directing the signal from AP $l$ to photodiode $n$ of user $u$, is given by
\vspace{-1mm}
\begin{equation}
\label{eq:ChannelModelNLoSRIS}
\hspace{-3mm} \begin{aligned}
& H_{l,k, \rm{ORIS}}^{[n,u]} {=} \\
 &
\begin {cases}
\hspace{-1mm}{{\hat{r}}  {\frac{{\left( {m + 1} \right) \cdot {A_{{\rm{PD}}}}}}{{2\pi {{\left( {{d_{l}^{[k]}} + {d_{k}^{[u]}}} \right)}^2}}}{{\cos }^m}{\left( {{\phi_{l}^{[k]}}} \right)}{\cos} \left( {{\varphi_{k}^{[n,u]}}} \right)}} & { 0 {\le} {\varphi_{k}^{[n,u]}} {\le} {\Psi}} \\
\hspace{-1mm}0 & \rm{otherwise}, \\
\end{cases}
\end{aligned}
\end{equation}
where $d_{l}^{[k]}$ and $d_{k}^{[u]}$ are the Euclidean distances from AP $l$ and ORIS element $k$, and from  ORIS element $k$ to  user $u$, respectively. The irradiance angle from AP $l$ to ORIS element $k$ is denoted by $\phi_{l}^{[k]}$, and $\varphi_{k}^{[n,u]}$ is the incidence angle from ORIS element $k$ to photodiode $n$ of user $u$, as represented in Fig.\,\ref{fig:ScenarioNLoS}. The reflection coefficient of an ORIS element is denoted by $\hat{r}$.

The NLoS channel gain produced by a wall grid element $w$ coming from AP $l$ to photodiode $n$ of user $u$ is given by
\vspace{-1mm}
\begin{equation}
\label{eq:ChannelModelNLoSdiff}
\hspace{-3mm} \begin{aligned}
 & H_{l,w, \rm{wall}}^{[n,u]} {=} \\
 &  \begin {cases} \hspace{-1mm}
{{\tilde{r}} \frac{{\left( {m + 1} \right)  {A_{{\rm{PD}}}}}}{{2{\pi}\left(\hspace{-0.5mm}d_{l}^{[w]}\hspace{-0.5mm}\right)^2\hspace{-0.5mm}\left(\hspace{-0.5mm}d_{w}^{[u]}\hspace{-0.5mm}\right)^2}}\hspace{-1mm}{A_k}{{\cos }^m}{\left( \hspace{-0.5mm} {{\phi_{l}^{[w]}}} \hspace{-0.5mm}\right)}{\cos \hspace{-0.5mm}\left(\hspace{-0.5mm} {{\varphi_{l}^{[w]}}} \hspace{-0.5mm}\right)}} { {\cos \hspace{-0.5mm} {\left( \hspace{-0.5mm} {{\phi_{w}^{[u]}}} \hspace{-0.5mm}\right)}}{\cos \hspace{-0.5mm} {\left( \hspace{-0.5mm}{{\varphi_{w}^{[n,u]}}} \hspace{-0.5mm}\right)}}} \\& {\hspace{-3cm} 0 {\le} {\varphi_{w}^{[n,u]}} {\le} {\Psi}} \\
\hspace{-1mm} 0 & \hspace{-3cm}\rm{otherwise}, \\ 
\end{cases}
\end{aligned}\quad
\end{equation}
where $\tilde{r}$ is the reflection coefficient of a wall surface, $\phi_{l}^{[w]}$ is the irradiance angle from AP $l$ to wall element $w$, $\varphi_{l}^{[w]}$ is the incidence angle onto wall element $w$ coming from AP $l$, $\phi_{w}^{[u]}$ is the irradiance angle from wall element $w$ to user $u$, and $\varphi_{w}^{[n,u]}$ is the incidence angle from wall element $w$ to photodiode $n$ of user $u$.

Due to Snell's law, each ORIS element can at most redirect the light from a single AP $l$ to a photodiode $n$ of a specific user $u$. Consequently, a resource allocation strategy is required to assign each ORIS element $k$ to a particular AP $l$ and photodiode $n$ of user $u$.  This assignment is represented by a Boolean variable defined as
\vspace{-1mm}
\begin{equation}
\beta_{l,k}^{[n,u]} = \begin {cases}
1, & \text{if $k$ is associated with AP $l$ }\\ & \text{and photodiode $n$ of user $u$} \\
0, & \text{otherwise}. \\
\end{cases}
\end{equation}

The NLoS channel gain from AP $l$ to photodiode $n$ of user $u$ is thus formulated as
\vspace{-1mm}
\begin{IEEEeqnarray}{rCl}
\hspace{-4mm} H_{l,\rm{NLoS}}^{{[n,u]}}\hspace{-1mm}\left( \hspace{-0.5mm}{{\beta_{l,k}^{[n,u]}}} \hspace{-0.5mm}\right) & {=} & \hspace{-0.5mm} \sum_{w} \hspace{-1mm} I_{l,w}^{[u]} H_{l,w, \rm{wall}}^{[n,u]}  {+} \hspace{-1mm} \sum_{k} \hspace{-1mm} I_{l,k}^{[u]} H_{l,k, \rm{ORIS}}^{[n,u]} {\beta_{l,k}^{[n,u]}},
\label{eq:ChannelModelNLoS}
\end{IEEEeqnarray}
where the binary indicators $I_{l,w}^{[u]}$ and $I_{l,k}^{[u]}$ determine the blockage state of the NLoS connection from AP $l$ to user $u$, via reflecting elements $w$ and $k$, respectively. These variables are set to 0 if the path is obstructed, e.g., by the user's body or other individuals, and to 1 if the connection is not obstructed. Consequently, the association $\beta_{l,k}^{[n,u]}$ between the AP, the ORIS element, and the photodiodes of users plays a crucial role in determining the system's overall effectiveness.

The overall VLC channel gain from AP $l$ to the photodiode $n$ of user $u$ can be formulated as
\begin{IEEEeqnarray}{rCl}
H_{l}^{[n,u]}\left( {{\beta_{l,k}^{[n,u]}}} \right) & {=} & H_{l, \rm{LoS}}^{{[n,u]}} + H_{l,\rm{NLoS}}^{{[n,u]}}\left( {{\beta_{l,k}^{[n,u]}}} \right).
\label{eq:ChannelModelTotal}
\end{IEEEeqnarray}

Without loss of generality, we implement a DC optical orthogonal frequency division multiplexing (DCO-OFDM) scheme composed of $N_{\rm sc}$ subcarriers. The total optical power per AP is $P_{\rm tot} = P_{\rm sc} \times \sqrt{N_{\rm sc}-2}$ since the first and $N_{\rm sc}/2$-th subcarriers do not transmit data, where $P_{\rm sc}$ is the optical power per subcarrier. To eliminate the possibility of interference, we consider that subcarriers are uniquely assigned to users, with the number of subcarriers at least as large as the number of users. Let us further assume that all APs transmit the same data to contribute to all users. Therefore, the SNR of photodiode $n$ of user $u$ can be formulated as
\begin{IEEEeqnarray}{l}
    {\gamma}^{[n,u]}({\boldsymbol{\beta}})  = 
    \frac{{{{\left( {{{\rho\cdot P_{\rm sc}}\sum\limits_l H_{l}^{[n,u]}\left( {{\beta_{l,k}^{[n,u]}}} \right)} } \right)}^2}}}{{{N_0}B}}, 
\label{eq:SNR}
\end{IEEEeqnarray}
where $\boldsymbol{\beta}$ is a matrix containing all $\beta_{l,k}^{[n,u]},\,\forall l,k,n,u$ values. The parameter $\rho$ is the photodiode responsivity, $N_0$ is the power spectral density of the additive white Gaussian noise (AWGN) at the receiver, mainly produced by shot and thermal noise~\cite{VLCLoSModel}, and $B$ is the communication bandwidth.

The rate produced by photodiode $n$ of user $u$, measured in bit/s/Hz can be formulated as 
\begin{equation}
\label{eq:UserRate}
\eta^{[n,u]}({\boldsymbol{\beta}}) = \log_2\left(1+\frac{\rm{exp(1)}}{2\pi}\cdot\gamma^{[n,u]}({\boldsymbol{\beta}})\right).
\end{equation}
Note that we have not invoked the Shannon capacity, but a tight lower bound of the capacity in an intensity modulation and direct detection system~\cite{CapacityVLC}.


\section{Proposed optimization problem}\label{sec:OptimizationProblem}

We propose an optimization problem to maximize the minimum SNR in an ORIS-assisted VLC environment. We aim to associate APs, ORIS and photodiodes of users optimally, which will allow us to understand to what extent ORIS elements and ADRs can complement each other to enhance VLC performance. First, due to the select-best combining technique as the photodiode selection technique, the SNR of user $u$ can be reformulated as 
\begin{equation}
\label{eq:SNRUseru}
{\gamma}^{[u]}({\boldsymbol{\beta}}) = \max\limits_n \gamma^{[n,u]}(\boldsymbol{\beta}),
\end{equation}
which considers only the SNR produced by the best-performing photodiode of the ADR of user $u$. Starting from the \texttt{maxmin} utility function, we can formulate the problem as fairness maximizing,
\begin{IEEEeqnarray}{l}
\label{MINOUT_MultiUser2}
\boldsymbol{\beta^*} = \mathop {\mathop {\argmax } \limits_{{\boldsymbol{\beta}}} \quad \min\limits_u \gamma'^{[u]}(\boldsymbol{\beta})}\limits\\
\begin{array}{l}
{\rm{subject\,to}}\\
{\rm{C1:  }}\,\sum\limits_l\sum\limits_u\sum\limits_n {{\beta _{l,k}^{[n,u]}} \le {1}}, \forall \, k \\
{\rm{C2:  }}\,{\beta _{l,k}^{[n,u]}} \in \{ 0,1\} ,{\rm{  }}\, \forall \, l,k,u,n\\
\end{array} \IEEEnonumber
\end{IEEEeqnarray}
For mathematical simplicity, we utilize the optical SNR defined as $\gamma'^{[u]}(\boldsymbol{\beta})  = \sqrt{\gamma^{[u]}(\boldsymbol{\beta})}$ in our objective function, keeping $\gamma'^{[u]}(\boldsymbol{\beta})$ as a linear function of the optimization variables $\boldsymbol{\beta}$. Constraint C1 restricts the allocation of every ORIS element to a single AP, user, and photodiode, and constraint C2 defines the optimization variables. Note that, although the user and photodiodes are considered to be placed in the same location, the orientation of each photodiode is different, which leads to different contributions from each ORIS element and, consequently, to different communication performance depending on the photodiode selected. 

We revise the objective function by introducing a new decision variable $\gamma_{\rm min}'=\min\limits_u \gamma'^{[u]}$, which is maximized in the optimization problem and leads to a new constraint $\gamma'_{\rm{min}}\leq \gamma'^{[u]}, \forall \, u$. In addition, we introduce a new variable $\boldsymbol{P}= \begin{bmatrix}\boldsymbol{p}_{1}\dots \boldsymbol{p}_{U} \end{bmatrix}$, which is a matrix of dimensions $N \times U$, where $\boldsymbol{p}_{u} = \begin{bmatrix}p_{u,1}, \dots, p_{u,N} \end{bmatrix}^T$ is a vector containing Boolean variables that take the value $p_{u,n}=1$ when the $n$-th photodiode of the $u$-th user is the one obtaining the best communication performance for user $u$, and $p_{u,n}=0$ otherwise. Then, the new optimization problem can be formulated as
\begin{IEEEeqnarray}{l}
\label{MINOUT_MultiUser1}
[\boldsymbol{\beta^*}, \gamma_{\rm min}'^*, \gamma'^{[u]*}, \boldsymbol{P^*}] = \mathop {\mathop {\argmax } \limits_{{\boldsymbol{\beta}, \gamma_{\rm min}', \gamma'^{[u]}, \boldsymbol{P}}} \quad \gamma_{\rm min}'}\limits\\
\begin{array}{l}
{\rm{subject\,to\,C1,\,C2,}}\\
{\rm{C3:  }}\,\gamma'_{\rm{min}}\leq \gamma'^{[u]}, \forall \, u \\
{\rm{C4:  }}\,\gamma'^{[u]}\geq \gamma'^{[n,u]}(\boldsymbol{\beta}), \forall \, u,n \\
{\rm{C5:  }}\,\gamma'^{[u]}\leq \gamma'^{[n,u]}(\boldsymbol{\beta})+M\cdot(1-p_{u,n}), \forall \, u,n \\
{\rm{C6:  }}\,\sum\limits_{n=1}^N\, p_{u,n} = 1\, \forall \, u\\
{\rm{C7:  }}\,p_{u,n} \in \{0,1\}\, \forall \, u.\\
\end{array} \IEEEnonumber
\end{IEEEeqnarray}
Constraint C3 defines $\gamma_{\rm min}'=\min\limits_u \gamma'^{[u]}$ as explained before. Constraint C4 determines the lower bound for $\gamma'^{[u]}$ according to the select-best combining technique for every user. To avoid unbounded results, constraint C5 determines the upper bound for $\gamma'^{[u]}$ according to the select-best combining technique for all users. For this purpose, we follow the big-$M$ approach~\cite{BigMApproach}, where $M$ is considered as an upper bound of $\gamma'^{[u]}$ for every possible user $u$. Note that C5 is satisfied both when $n$ is selected as the photodiode offering the best performance ($p_{u,n}=1$), and when $n$ is not the photodiode offering the best performance ($p_{u,n}=0$) because $M$ is a very large value. Constraint C6 guarantees that only one photodiode per user $u$ is selected as the one offering the best performance. Finally, constraint C7 defines the variables $p_{u,n}$ to optimize. The problem formulated is an Integer Linear Programming (ILP) problem, which is convex by definition and can be solved by a solver such as Gurobi and CVX in Matlab~\cite{cvx}.

\setlength{\tabcolsep}{4pt}
\begin{table}[!t]
\caption{Simulation parameters.}
\label{tab:SystemParameters}
\centering
{\footnotesize
\begin{tabular}{llcr}
\hline
\hline
Notation & Parameter description & Value & Unit\\
\hline
- & AP height &  3 & [m]\\
- & User device height &  1 & [m]\\
$\phi_{1/2}$ & Half-power semi-angle of  the LED &  80 & [deg.]\\
$\tilde{r}$ & Reflection coefficient of wall & 0.4 & [-]\\
$\hat{r}$ & Reflection coefficient of ORIS & 0.95 & [-]\\
$\rho$ & Photodiode responsivity & 0.4 & [A/W]\\
$A_{\mathrm{PD}}$ & Photodiode physical area & 1 & [cm$^2$]\\
$\Psi$ & FoV semi-angle of each photodiode & [15, 45, 75] & [deg.]\\
$B$ & Bandwidth & 20 & [MHz]\\
$N_0$ & Power spectral density of the AWGN & $2.5{\cdot}10^{-20}$ & [W/Hz]\\
\hline
\hline
\end{tabular}}
\end{table}
\setlength{\tabcolsep}{6pt}

\begin{figure}[t]
\centering
\includegraphics[width=\columnwidth]{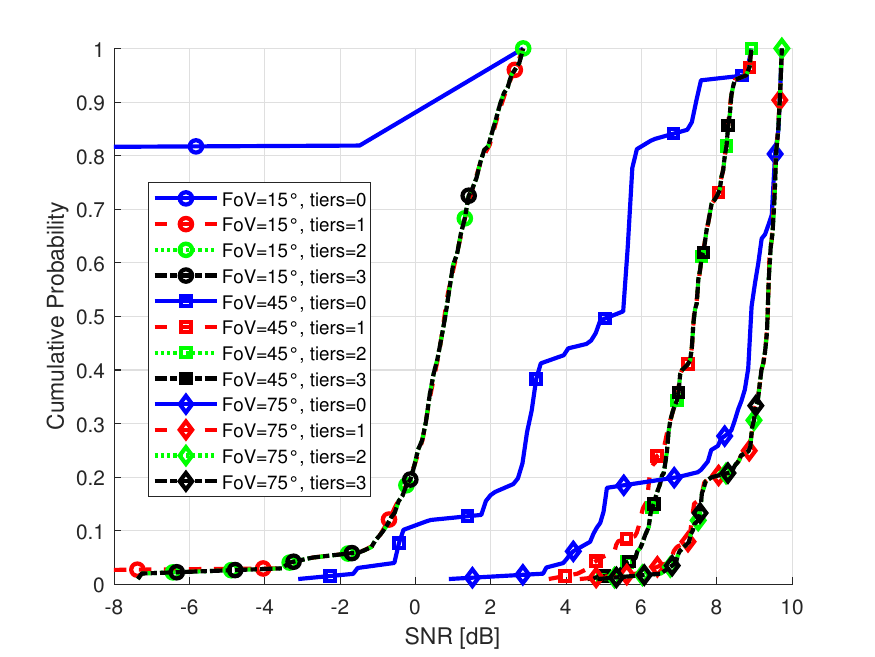}
        \caption{CDF of SNR without ORIS elements deployed; single user.}
\label{fig:CDF_SNR_NORIS}        
\vspace{-7mm}
\end{figure}

\begin{figure}[t]
\centering
\includegraphics[width=\columnwidth]{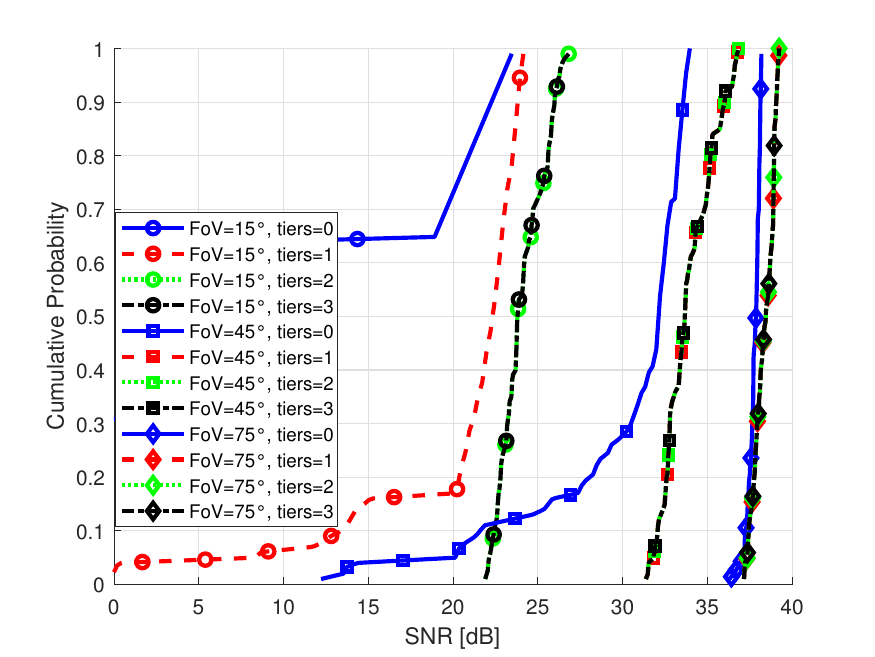}
        \caption{CDF of SNR with ORIS elements deployed; single user.}
\label{fig:CDF_SNR_ORIS}        
\vspace{-5mm}
\end{figure}

\begin{figure}[t]
\centering
\includegraphics[width=\columnwidth]{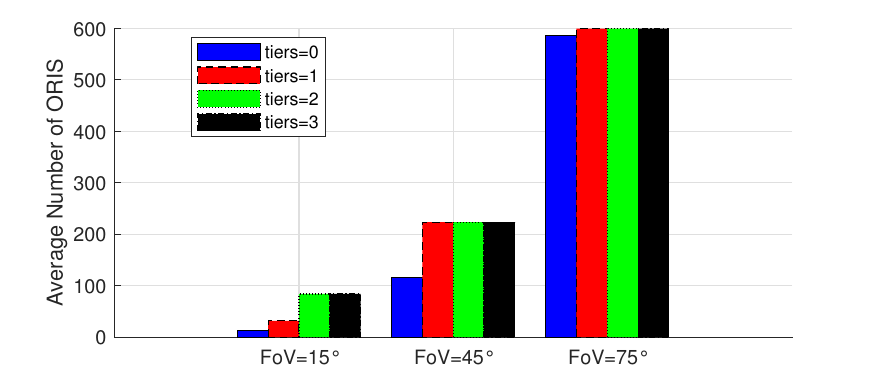}
        \caption{Average number of ORIS elements used for each configuration of FoV and ADR type; single user.}
        \vspace{-7mm}
\label{fig:NRIS}
\end{figure}

\begin{figure}[t]
     \centering
     \begin{subfigure}[t]{0.49\columnwidth}
         \centering
         \includegraphics[width=\textwidth]{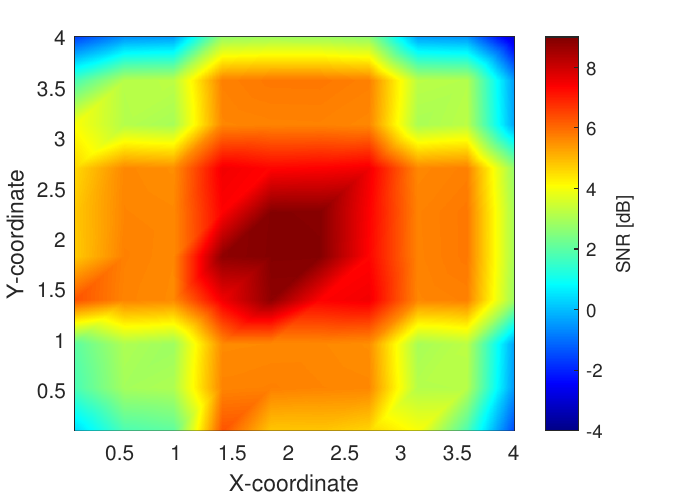}
         \caption{Photodiode}
         \label{fig:2D_PD_NORIS}
     \end{subfigure}
     \hfill
     \begin{subfigure}[t]{0.49\columnwidth}
         \centering
         \includegraphics[width=\textwidth]{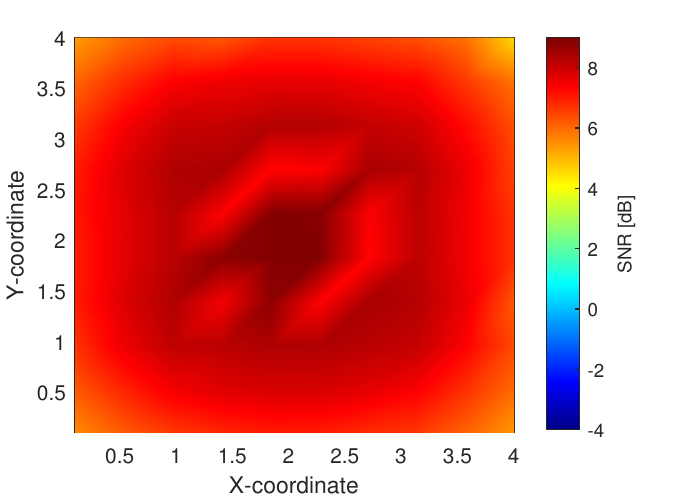}
         \caption{ADR}
         \label{fig:2D_ADR_NORIS}
     \end{subfigure}
        \caption{Heat map of SNR in the room without ORIS elements deployed; single user.}        
        \label{fig:2D_NORIS}
        \vspace{-5mm}
\end{figure}

\begin{figure}[t]
     \centering
     \begin{subfigure}[t]{0.49\columnwidth}
         \centering
         \includegraphics[width=\textwidth]{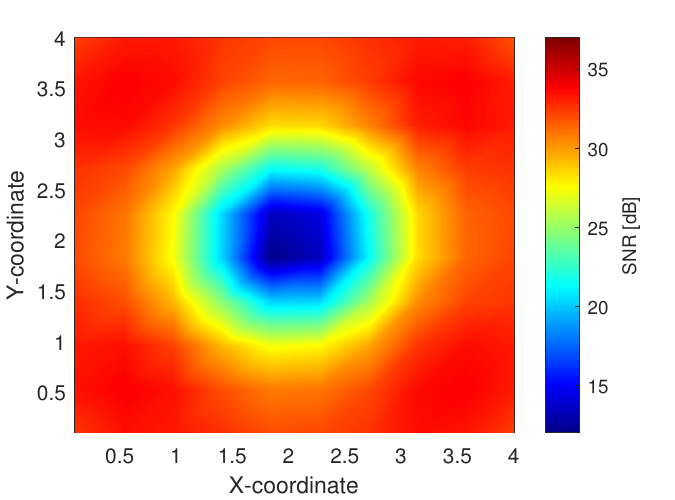}
         \caption{Photodiode}
         \label{fig:2D_PD_ORIS}
     \end{subfigure}
     \hfill
     \begin{subfigure}[t]{0.49\columnwidth}
         \centering
         \includegraphics[width=\textwidth]{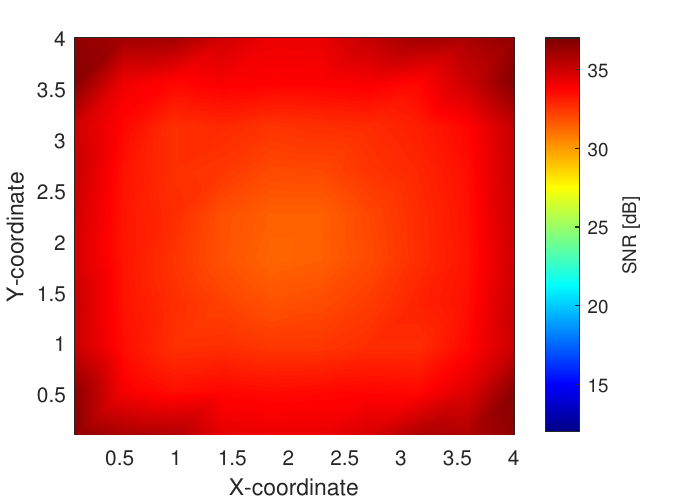}
         \caption{ADR}
         \label{fig:2D_ADR_ORIS}
     \end{subfigure}
        \caption{Heat map of SNR in the room with ORIS elements deployed; single user.}        
        \label{fig:2D_ORIS}
        \vspace{-7mm}
    \end{figure}

\section{Simulation results}\label{sec:Results}
In this section, we present a detailed analysis of the simulation results obtained for the VLC scenario tested. This scenario features a deployment of ORIS elements that occupies the upper one-third of every wall. We analyze a typical 4$\times$4$\times$3 meter empty room, where a total of $L=4$ APs are arranged symmetrically at coordinates [1,1], [1,3], [3,1], and [3,3] meters. The NLoS links, i.e., reflections, from all four walls, considering both the ORIS and the remaining wall surfaces, are incorporated into the simulation. The number of ORIS elements used per wall is $K=30\times 5=150$. The simulation parameters are summarized in Table\,\ref{tab:SystemParameters}. 

Each user body is modelled by a cylinder of height 1.75\,m and radius 0.15\,m, with a receiving device separated at a distance of 0.3\,m from the body~\cite{CylinderBlockage}. The users are uniformly distributed along the room, each oriented at any horizontal angle following a uniform distribution $\mathcal{U}[0, 2\pi)$, and they can block their own and other users' LoS and NLoS links. 

Each user is provided with a receiver that may be either a single photodiode looking upwards (tier 0), or an ADR composed of multiple photodiodes arranged as detailed in Fig.\,\ref{fig:reconfigurable}. The elevation angle of photodiodes located at tier 1, tier 2, and tier 3 are $\theta=30^\circ$, $\theta=60^\circ$ and $\theta=90^\circ$, respectively, with respect to the vertical axis, and the azimuthal angle between consecutive photodiodes located at tier 1, tier 2, and tier 3 is $\alpha=60^\circ$, $\alpha=30^\circ$ and $\alpha=20^\circ$, respectively. This leads to a total number of photodiodes in the ADR of $N=7$, $N=19$, and $N=37$ in case of a tier-1 ADR, tier-2 ADR, and tier-3 ADR, respectively. 

First, we study the cumulative density function (CDF) of the SNR of a single user randomly located around the room according to a uniform distribution, considering that there is no blockage. Note that this is an upper bound of the SNR, as ideal conditions are guaranteed, i.e., no blockage and all resources in the case of an ORIS-assisted VLC scenario are dedicated to the same user. We compare the results obtained for all possible configurations: single photodiode (tiers=0) and ADR (tiers=1, 2, or 3), for three possible FoV values ($\Psi=15^\circ$, 45$^\circ$, 75$^\circ$). Fig.\,\ref{fig:CDF_SNR_NORIS} plots the results obtained when no ORIS element is deployed, whereas Fig.\,\ref{fig:CDF_SNR_ORIS} represents the results obtained when ORIS elements are deployed. We observe that the larger the FoV, the larger the SNR is, as further contributions from neighboring APs and from reflecting elements are received. The same phenomenon is observed with the number of tiers, as it increases the total FoV of the whole receiver. However, there is no big difference between ADRs with tiers 1, 2 or 3, which means that implementing an ADR with 1 tier, i.e., a total number of 7 photodiodes (1 in the middle and 6 in tier 1, as depicted in Fig.\,\ref{fig:reconfigurable}), suffices, this way maintaining a trade-off between receiver complexity and communication performance. For the remainder of the paper, we consider a single tier when implementing an ADR as a receiver. Providing a single photodiode to users (tiers=0) may lead to outages, as the SNR tends to very low values when narrow FoV values are considered.

When comparing results obtained without ORIS elements (Fig.\,\ref{fig:CDF_SNR_NORIS}) and with ORIS elements (Fig.\,\ref{fig:CDF_SNR_ORIS}), we see that deploying ORIS elements may increase the SNR of a user up to around 30\,dB, which can be seen as a diversity gain because multiple copies of the same signal are received, and they are coming from many different paths. These results highlight the communication gain produced by ORIS deployment. Using ORIS also increases the robustness of the VLC channel against potential link blockages, as discussed below.

\begin{figure}[t]
\centering
\includegraphics[width=\columnwidth]{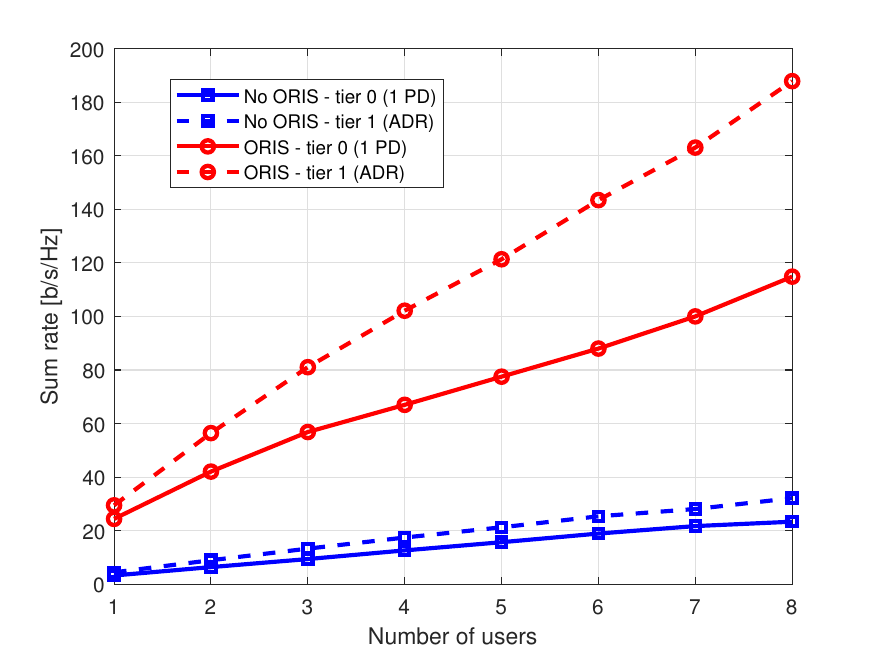}
        \caption{Sum rate considering blocking elements.}
\label{fig:SumThroughput}      
\vspace{-5mm}
\end{figure}

The average number of ORIS elements associated with the user for each FoV and ADR type configuration is depicted in Fig.\,\ref{fig:NRIS}. 
Note that, unless the FoV equals 15$^\circ$, all ADR configurations exploit the ORIS deployment equally, i.e., the number of ORIS elements exploited is the same for all ADR configurations. This confirms that providing users with an ADR assuming a single tier (tiers=1) suffices to exploit an ORIS deployment. Also, we observe that the larger the FoV, the larger is the number of ORIS elements used, as signals coming from ORIS elements that produce larger incidence angles may be received.

To analyze the room locations where the ORIS and ADR have a larger impact on the communication performance, we now study the SNR distribution around the room. We consider a typical FoV value of 45$^\circ$. For this purpose, Fig.\,\ref{fig:2D_NORIS} plots in a 2D heat map the received SNR without ORIS deployment, where we can analyze the ADR contribution. Note that the locations that benefit more from an ADR deployment are the ones in the center of the room, i.e., where the user may receive signals coming from multiple APs. The diversity produced by the ADR deployment makes the SNR distribution much more homogeneous, which leads to a network with a large communication fairness, as observed in Fig.\,\ref{fig:2D_ADR_NORIS}. 

To determine the ORIS contribution, the SNR distribution assuming an ORIS deployment is shown in Fig.\,\ref{fig:2D_ORIS} (Note the scale difference between the heat maps in Figs.\,\ref{fig:2D_NORIS} and \ref{fig:2D_ORIS}.)  When comparing Fig.\,\ref{fig:2D_PD_NORIS} and Fig.\,\ref{fig:2D_PD_ORIS}, we observe that the ORIS deployment makes the users located at the room edges the ones with a higher communication performance. There is an SNR difference of around 20\,dB between a user located at the room center and at the room edge, which shows the ORIS capability to strengthen the received signal at the user position. When both ORIS and ADR are deployed (Fig.\,\ref{fig:2D_ADR_ORIS}), the whole room area is provided with a very similar SNR, and it is much more powerful than the network provided without ORIS (Fig.\,\ref{fig:2D_ADR_NORIS}).

We finally evaluate the sum rate performance when multiple users are deployed. We consider a realistic scenario with self-blocking and blockage among different users, both with LoS and NLoS links. Again, we consider a FoV equal to 45$^\circ$ and 1 tier for the ADR case. Results are depicted in Fig.\,\ref{fig:SumThroughput}. Note that, the larger the number of users deployed, the higher the sum rate. Besides, when ORIS elements are deployed, larger sum rates are obtained due to an SNR increase as observed in previous results. The use of ADRs is more determinant when deployed together with ORIS elements, which shows their complementarity. In addition, when the number of users deployed increases, the gain obtained by the ADR is much more impactful compared to the gain obtained by a single photodiode. In fact, the sum rate when both ORIS and ADR are employed scales with the number of users. The reason is that the number of blockages produced when deploying more users increases, but providing secondary paths enabled by ADRs and ORIS elements overcomes this issue.

\section{Conclusion} \label{sec:Conclusions}
This paper investigates the contribution of ORIS elements and ADRs to improve the VLC performance. We formulated an optimization problem to optimally associate ORIS elements to users and APs, with the objective of maximizing the minimum SNR among multiple users, when users can block their own communication links and the ones corresponding to other users, both LoS and NLoS. We implemented a select-best combining technique to choose the photodiode for each user. We show that ADR and ORIS can complement each other to provide a large diversity gain to user receivers, as well as to provide secondary communication links that make the VLC system more robust against blockages. This translates into SNR gains of more than 30\,dB when deploying ORIS elements, and larger communication fairness and sum rates when providing users with ADRs, which make them further exploit the NLoS links coming from ORIS elements.

\bibliography{./references}
\bibliographystyle{IEEEtran}

\end{document}